\documentclass[review]{elsarticle}
\usepackage{xcolor}
\usepackage{subcaption}
\usepackage{lineno,hyperref}
\modulolinenumbers[5]
\usepackage[colorinlistoftodos]{todonotes}
\usepackage{multirow}
\newcommand{\specialcell}[2][c]{%
  \begin{tabular}[#1]{@{}c@{}}#2\end{tabular}}

\journal{Information Processing and Management}

\begin{document}

\begin{frontmatter}

\title{Changing Views: Persuasion Modeling and Argument Extraction from Online Discussions}

\author{Subhabrata Dutta, Dipankar Das}
\address{Jadavpur University, Kolkata, India}
\ead{\{subha0009,dipankar.dipnil2005\}@gmail.com}

\author{Tanmoy Chakraborty}
\address{IIIT-Delhi, New Delhi, India}
\ead{tanmoy@iiitd.ac.in}




\begin{abstract}
Persuasion and argumentation are possibly among the most complex examples
of the interplay between multiple human subjects.   With the advent of the Internet,
online forums provide wide platforms for people to share their opinions and
reasonings around various diverse topics.   In this work, we attempt to model
persuasive interaction between users on Reddit, a popular online discussion forum.  We
propose a deep LSTM model to classify whether a conversation leads to a successful persuasion or not, and use this model to  predict whether
a certain chain of arguments can lead to persuasion.   While learning persuasion dynamics, our model tends to identify argument facets implicitly, using an
attention mechanism.  We also propose a semi-supervised approach to extract
argumentative components from discussion threads.  Both these models provide
useful insight into how people engage in argumentation on online discussion forums.
\end{abstract}


\end{frontmatter}

\section{Introduction}
\label{sec:intro}
Argumentation, to put simply, is posing some claim around a concept along with one or more premises supporting that claim. Persuasion is an application of argumentation, where two or more people exchange contending arguments to convince other participants to change their initial positions. Argumentation and persuasion in general play crucial roles in versatile social activity, from legal dialogues to scientific reasoning and many more. \textcolor{black}{With social media platforms, online interaction on almost every aspect of life for a modern-day person is now a concrete possibility, that too among a huge sect of people. As discussed by Fogg \cite{fogg2008mass}, debate, argumentation and persuasion have also found their ways to this world connected by web. Studying such complex human interactions at scale would be useful in different applications -- automated learning strategy, social network security, political-sociological analysis, marketing with a more minute knowledge that a customer needs, etc. }

Although argumentation has somewhat objective qualifications to test for, persuasion is more of a subjective issue. That is, a third observer can decide to an extent whether two participants are arguing or not, and judge the quality of arguments. Whether someone will be persuaded by someone else's arguments surely depends on this \textit{quality of arguments}; but the final decision is to be made by the one being persuaded. One can readily observe conversations over a real and virtual medium where conversations do not lead to persuasion, though one participant is clearly out of arguments and simply holds onto some claims/beliefs devoid of logic. 
\par Modeling argumentation from natural language texts has gained much attention among researchers recently. As Lippi and Torroni  \cite{lippi2016argumentation} described, mining arguments from raw texts can be though of as a two-stage process -- \textit{argumentative component extraction} and \textit{argument structure identification}. The first stage can further be divided into extracting argumentative sentences from raw text and identifying argument component boundaries in those sentences. There is no single, universally consensus argumentation model. Toulmin's model  \cite{stephentoulmin2003} received much usefulness over the last decade. Some other notable models were given by Farley and Freeman \cite{farley1995burden}, 
Dung \cite{dung1995acceptability}. A rather simplified definition of argument components was introduced by Aharoni et al. \cite{aharoni2014benchmark} -- they identified \textit{claims} as text segments, expressing a clear stance towards some concept, and \textit{premises} supporting that claim. In this work, we will follow this notion of argumentation.
\par  But not all platforms provide space for discussions which can be studied to model persuasion dynamics. Microblogging sites like Twitter are often used as campaign platforms -- with restrictions on the size of the text, primary focus of a user is to showcase her/his opinion, rather than engaging in some constructive debate. Platforms like Reddit, Livejournal, CreateDebate, etc. are better suitable candidates to study persuasion and argumentation in online conversations as they are designed for promoting debate-like discussions.

\textcolor{black}{However, the study of persuasion and argumentation in online discussions bears its own challenges. To decide whether a person is persuaded or not is very difficult sometimes from the point of view of an outsider. For example, it is possible that a person did not comment after being persuaded by an argument. None other than the subject can decide in that case whether s/he was persuaded successfully or not. Similar difficulty can arise if multiple users are engaged to persuade someone. There is no way to directly infer which chain of arguments led to successful persuasion, unless the persuaded person herself/himself expresses.}

\textcolor{black}{The difficulty of finding annotated data accompanies the task of argument mining too. Argumentation models, as discussed earlier, may facilitate objective annotation of the data; but it is a complicated linguistic task. As Habernal and Gurevych \cite{habernal2017argumentation} reported, most of the available datasets for argumentation in user-generated web discourse have a low inter-annotator agreement and are typical of small size. This makes supervised training for argumentation mining a challenging task and poses a need for unsupervised/weakly-supervised algorithms to employ. 
}

\begin{figure}[!t]
    \centering
    \includegraphics[width=0.9\textwidth]{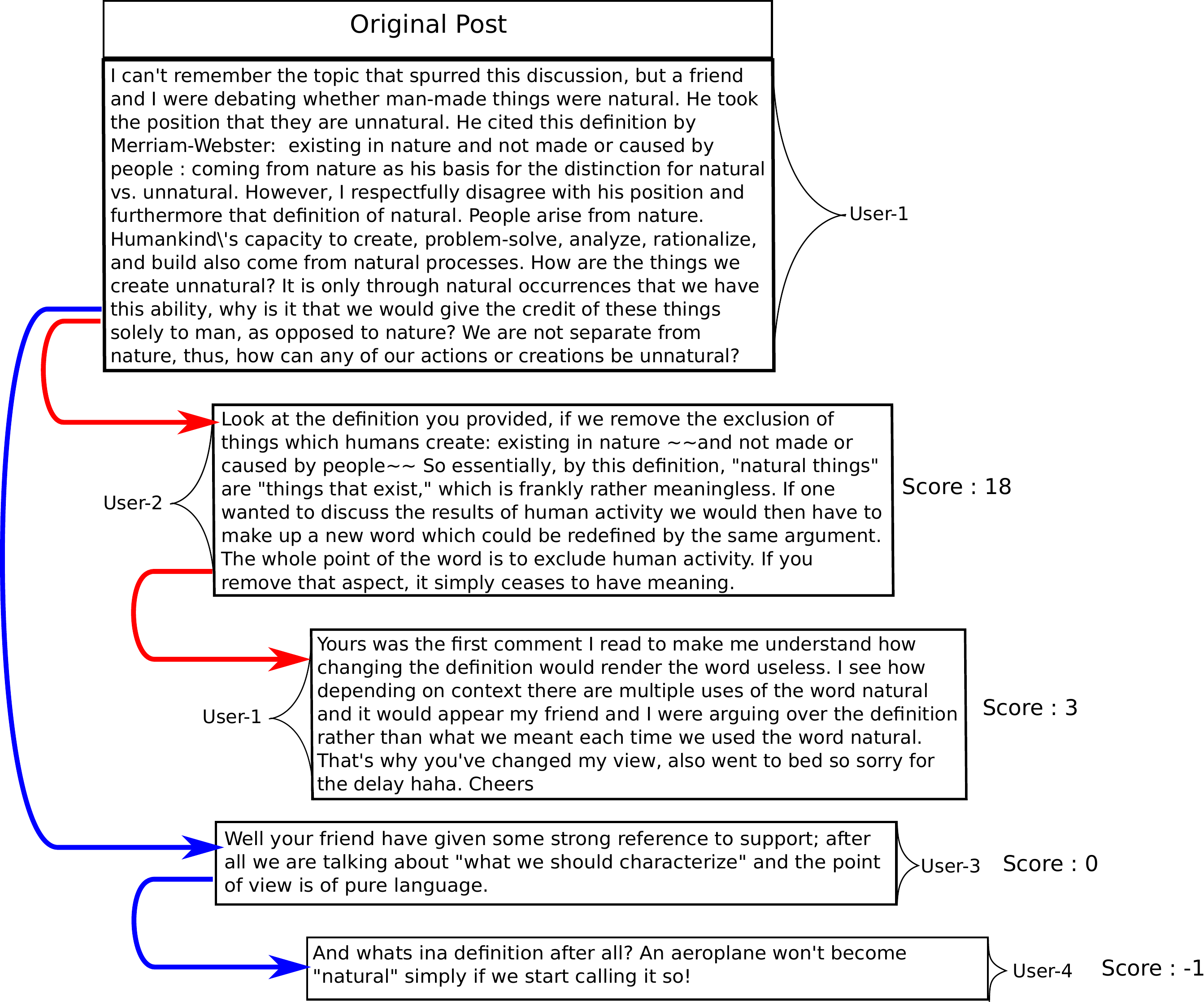}
    \caption{(Color online) An example discussion thread from Reddit CMV. Singly colored arrows represent chains -- red one identifying a successful persuasion. Score is the karma score acquired by the comment.}
    \label{fig:example_thread}
\end{figure}

\textcolor{black}{To tackle the labeling challenge for persuasion modeling, it would be best if the discussion platform itself provides an option for users to acknowledge whether they have been successfully persuaded or not.} A sub-community of Reddit, namely \verb|/r/ChangeMyView| (aka CMV) is one such candidate \cite{tan2016winning}. A user (henceforth named \textit{Original Poster} or OP) posts his/her views towards a certain topic, and people engage in argumentation to change that stance. If some chains of arguments become successful to change OP's view, a \textit{delta} ($\Delta$) is awarded by OP. A comment awarded with a $\Delta$ means, the discussion subtree rooted at that comment was successful to change OP's view. As the forum is highly moderated, comments are usually argumentative, containing reasoning rather than sarcastic bullying. Also, each comment is associated with a `\textit{Karma Score}' given by users denoting its argumentation quality. Fig.~\ref{fig:example_thread} is an example discussion thread from CMV subreddit.

\par \textcolor{black}{Though persuasion identification and argument mining may seem different linguistic tasks with their own merits and the vast existing body of knowledge, in the conversation arena, these two problems are closely related as we have already noted. As Biran et al. \cite{biran2012detecting} suggested, the identification of influencers in written dialogues (which is a persuasion modeling task) needs argument structure understanding as its subtask. The algorithms we propose in this work may further be integrated for a better understanding of both. Also, as our findings indicate, an end-to-end deep learning model performs argument mining implicitly to detect persuasion in conversations.} We attempt to tackle four related problems together: (i) detection and prediction of persuasion in online discussions; (ii) ranking most influential comments posted, identified from users' perspective; (iii) automatic extraction of argumentative sentences as a subtask of supervised persuasion modeling; and (iv) identification of argument components (claim/premise) from online discussion threads. To deal with these problems, our contributions in this work are as follows:
\begin{itemize}
\item We propose an attention-based Hierarchical LSTM network to identify successful persuasion from CMV conversations. We use this model to predict whether a particular chain of comments is going to persuade the OP or not. This model jointly learns the binary classification of persuasion guided by $\Delta$-score and regression task of predicting the karma score for each comment.
\item We incorporate an attention mechanism in our model which weighs sentences in a comment accordingly; we use these weights to identify argumentative sentences. \textcolor{black}{To the best of our knowledge, this is the first attempt to train a neural network for a supervised task (persuasion prediction) and force it to learn an unsupervised task (argumentative sentence identification).}
\item We propose a novel algorithm to detect argument components from argumentative sentences. This algorithm exploits some linguistic rules to identify a subset of argument components and finds the closure using similarity measurements. We employ Dynamic Time Warping distance (\textcolor{black}{a measurement of geometric similarity between variable length time series data}) to compute the similarity between text segments taken as time series. \textcolor{black}{To the best of our knowledge, this is the first semi-supervised approach for argument mining from social media conversations.}
\end{itemize}
\par Rest of the article is organized as follows. We describe the research works in the field of persuasion modeling and argument mining from online discussions in Sec.~\ref{sec:related}.
 In Sec.~\ref{sec:persuasion_model}, we present our deep LSTM based model with attention mechanism for persuasion detection.
In Sec.~\ref{sec:arg_ext}, we discuss our proposed algorithm for argument component identification. Datasets and other experimental setups are discussed in Sec.~\ref{experiment}. We present all the results observed in Sec.~\ref{sec:obs}.
We present a brief discussion on the observations, error analysis and possible future work in Sec.~\ref{sec:discussions}

\section{Related Works}
\label{sec:related}
\textcolor{black}{We defined our primary problems as persuasion modeling and argument understanding, with the focused domain being social media conversations. In this section, we briefly review the existing body of knowledge regarding these two problems in general, with emphasis on the chosen domain.}
\par From a linguistic point of view, studies on persuasion have gained attention for a much longer period; works by Crismore et al. \cite{crismore1993metadiscourse}, Holtgraves and Lasky \cite{holtgraves1999linguistic} 
are few of the many. The study of persuasion in written dialogues started only after the boom of social media. Biran et al. \cite{biran2012detecting} studied the problem of influential user detection in online discussions. Their work was based on Wikipedia talkpage discussions \textcolor{black}{(88 discussion threads) and Livejournal discussions (245 discussion threads). They built a pipeline system for influencer detection -- it identifies claims, argumentation, persuasion attempts, and agreement of the subject step-by-step. They also used dialogue patterns for this task. The proposed system achieved F1 score of 0.59 for the Wikipedia talkpage and 0.74 for the Livejournal dataset using Support Vector Machine.}  Quercia et al. \cite{quercia2011mood} presented a study on language usage of an influencer over Twitter. All of these studies were based on datasets of much smaller size. Weiksner et al. \cite{weiksner2008six} studied different patterns of persuasion over online social networks. \textcolor{black}{They identified six different patterns of persuasion over Facebook; four of these depend on native characteristics of Facebook, whereas two patterns are identified external sources (real-world events, interactions in other platforms, etc.).} Guadagno and Cialdini \cite{guadagno2002online} reported how gender identity plays role in persuasion over online discussions. 

 Tan et al. \cite{tan2016winning} presented the first study on persuasion dynamics on large scale data. They crawled discussion threads from CMV subreddit for a two-year long time period. \textcolor{black}{They tackled a two-way prediction -- whether a user attempting to pursue the OP will be $\Delta$-awarded, and whether a given OP will change her/his position (stance malleability). They achieved an accuracy of 0.7 for the first task and 0.54 AUC score for the second.} Wei et al. \cite{wei2016post} used the same dataset to study ranking problem of persuasive comments (\textcolor{black}{ranking comments according to karma score}). Their important observation was that argumentative textual features perform better for this task compared to surface textual features.
\par As discussed earlier, studies in computational argumentation or argument mining gained attention among researchers very recently. Arguments from various types of texts such as formal legal texts \cite{palau2009argumentation,moens2007automatic}, essays  \cite{stab2014annotating,stab2014identifying}, social semantic web \cite{schneider2013review}, etc. have been attempted. We restrict ourselves to the studies on argument mining from online discussions only. Goudas et al. \cite{goudas2014argument} attempted to extract argument components from Greek social media discussions related to renewable energy. They formulated the task as a two-stage process: the detection of argumentative sentences from raw text and identification of claim/premise segments. Biran et al. \cite{biran2011identifying} presented a similar work, identifying claim/justification segments from Livejournal and Wikipedia talkpage discussions. Though their work was on English discussions, their dataset is not publicly available to test the performance of other algorithms. Habernal and Gurevych \cite{habernal2016makes} showed an empirical study on deciding the persuasive quality of arguments from \textit{createdebate.com} and \textit{convinceme.net}. Ghosh et al. \cite{ghosh2014analyzing} developed a corpus from the Technorati blog post comments. They annotated argument components as \textit{targets} and \textit{callouts} with agreement/disagreement relations between them.

\section{Persuasion Modeling}
\label{sec:persuasion_model}
We formulate the task of persuasion modeling as a sequence classification problem. Given a chain of comments $C = \{c_0,c_1,...,c_n\}$, the problem becomes a binary classification task of deciding whether that sequence of comments has been awarded a $\Delta$ or not.
\par Recurrent neural networks (RNN), particularly, Long Short Term Memory networks (LSTMs) \cite{hochreiter1997long} have achieved much success in recent times for sequence learning tasks. They use a separate memory cell to remember long term dependencies, which can be updated depending on current input. At each time step, an LSTM takes current input $x_t$ and previous memory cell state $c_{t-1}$ as input and computes output $o_t$ and current cell state $c_t$, following the equations below \begin{equation}
i_t = \sigma_i(x_tW_{xi}+h_{t-1}W_{hi}+b_i)
\end{equation}
\begin{equation}
f_t = \sigma_f(x_tW_{xf}+h_{t-1}W_{hf}+b_f)
\end{equation}
\begin{equation}
\label{celleq}
c_t = f_t\odot c_{t-1}+i_t\odot \sigma_c(x_tW_{xc}+h_{t-1}W_{hc}+b_c)
\end{equation}
\begin{equation}
o_t = \sigma_o(x_tW_{xo}+h_{t-1}W_{ho}+b_o)
\end{equation}
\begin{equation}
h_t = o_t\odot \sigma_h(c_t)
\end{equation}
where $x_t$ is the input vector, $f_t$ is the forget gate activation vector, $i_t$ is the input gate activation vector, $o_t$ is the output gate activation vector, $h_t$ is the output vector of LSTM, $c_t$ is the cell state, and $W$ and $b$ are weight and bias matrices, respectively.
\begin{figure}[h]
\centering
\includegraphics[width=0.8\textwidth]{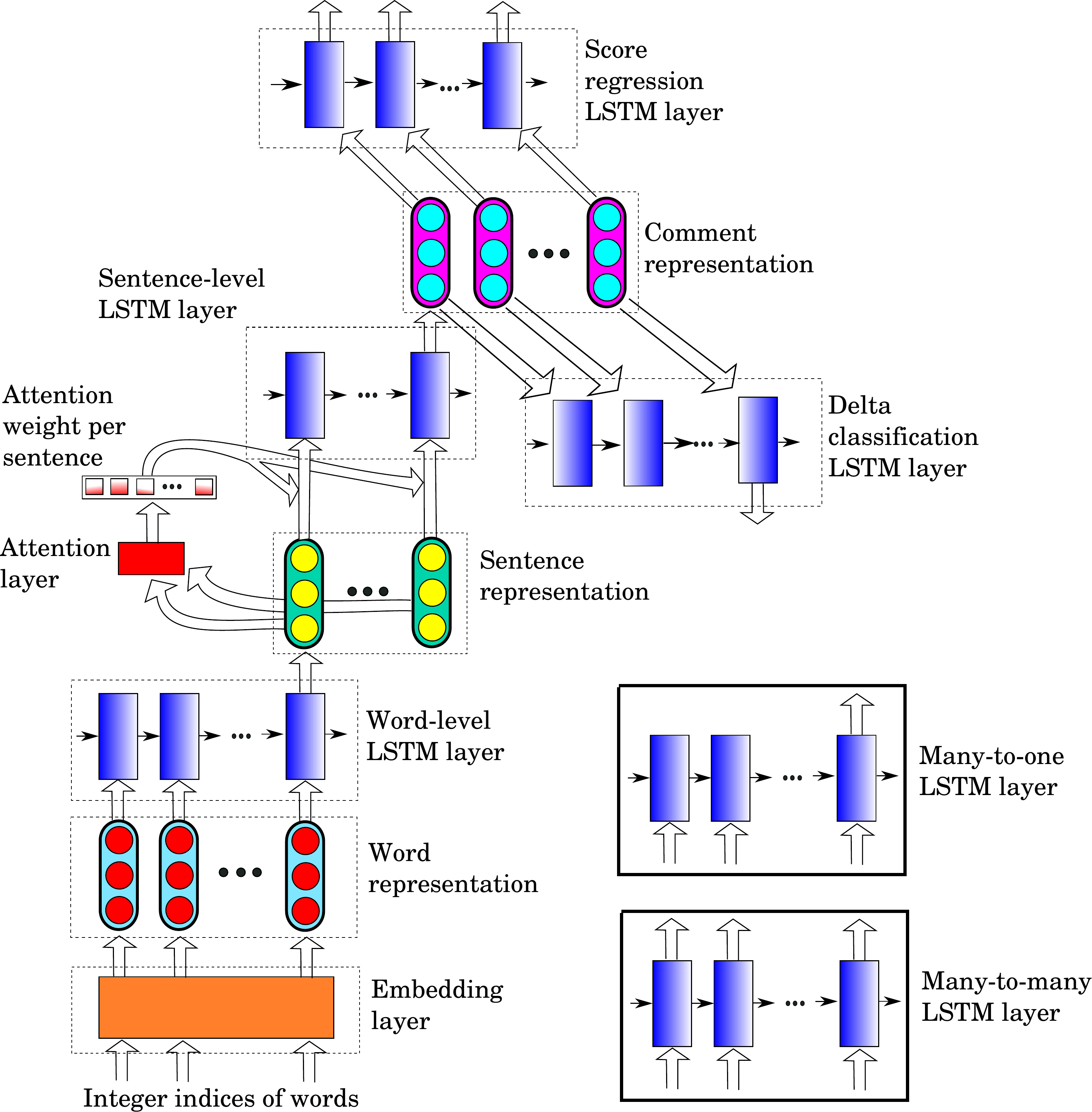}
\caption{(Color online) Persuasion modeling architecture using hierarchical LSTM with attention; hidden states of the $\Delta$-classification and score regression LSTM layers are initialized from another LSTM layer which takes words of the first comment (OP) as input (not shown here to maintain clarity); score regression layer maps every comment to a score (many-to-many), whereas $\Delta$-classification maps a chain of comments to a binary value representing successful/failed persuasion.}
\label{fig:persuasion_model}
\end{figure}
\par We use a hierarchical stacking of LSTM layers in our persuasion classification model. A single chain of comments is broken into two inputs -- the OP comment, $c_0$, and rest of the chain $\{c_0,c_1,...,c_n\}$. Each comment $c_i$ of the second input is delivered as a sequence of sentences $\{s_{0}^{i},s_{1}^{i}, ..., s_{m}^{i}\}$; where every sentence $s_{j}^{i}$ is a sequence of words $\{w_{0j}^{i},w_{1j}^{i}, ..., w_{lj}^{i}\}$. As depicted in Fig.~\ref{fig:persuasion_model}, the very first layer of our model takes each sentence as a sequence of words and constructs a vector representation for each sentence. Next, the sentence-level LSTM layer computes a sequence of sentence representations from output of previous layer. If we denote word-level and sentence-level LSTM layers as black-box functions $\mathcal{WL}$ and $\mathcal{SL}$ respectively, then previous two computations become
\begin{equation}
S = \mathcal{WL}(W)
\end{equation}
\begin{equation}
S' = \mathcal{SL}(S)
\end{equation}
Once per sentence representation is computed from the previous two layers, the attention mechanism attempts to learn the weights of each sentence in a comment according to their content and builds a comment representation. It computes a weight distribution $P = [p_0, p_1, ..., p_m]$ over all the sentences from $S' = [s'_0, s'_1, ..., s'_m]$ as follows:
\begin{equation}
p'_i = tanh(\mathbf{W}[i]\cdot s'_i + \mathbf{B}[i])
\end{equation}
\begin{equation}
p_i = \frac{e^{p'_i}}{e^{p'_i}+1}
\end{equation}
where $\mathbf{W}$ and $\mathbf{B}$ are matrices to be learned by the attention layer. Each sentence representation $s_i$ is then multiplied by $p_i$ and summed to get the comment representation $c_j$. 
\par We use the comment representation computed for a two-way learning task. An LSTM layer (henceforth named as {\em score regression layer}) maps the sequence of comment representations to a sequence of real-valued scores (collected from the karma scores given to each comment by Reddit CMV community). 
Another LSTM layer, namely $\Delta${\em -classification layer}, maps the comment sequence to a single value indicating a binary classification -- whether a $\Delta$ is awarded or not. Initial cell states of these two layers are initialized by the hidden state of a parallel LSTM layer which takes the OP comment as input. The idea is that  this LSTM will ``memorize'' useful parts of the OP comment, supply them to the $\Delta$ classification layer and score regression layer. 
\par This model is trained to minimize two objective functions $\mathit{l_1}$ and $\mathit{l_2}$ jointly due to two different outputs. 
\begin{equation}
\mathit{l_1} = \frac{\sum_{j=1}^{n} (s_j-\hat{s_j})^2}{n}
\end{equation}
\begin{equation}
\label{loss_2}
\mathit{l_2} = -(y\log p + (1-y)\log (1-p))
\end{equation}
where $s_j$ and $\hat{s_j}$ are the predicted and true scores of the j-th comment respectively; $y\in \{0,1\}$ denotes whether the chain was awarded $\Delta$, and $p\in[0,1]$ is the predicted probability of being $\Delta$ awarded. As scores are auxiliary measures to identify a comment's potential to be persuasive, we set a loss weight ratio 0.3:1 for $\mathit{l_1}$ and $\mathit{l_2}$, i.e., the final classification task is more focused by our model.

\section{Semi-supervised Argument Component Detection}
\label{sec:arg_ext}
In Sec.~\ref{sec:intro}, we discussed the stages of argument mining; the very first step being the separation of argumentative sentences from non-argumentative ones. As Lawrence et al. \cite{Lawrence2014-LAWMAF-2} suggested, the presence of topic-relevant words and phrases in a sentence is a good indicator of the sentence being a candidate. We compute TF-IDF scores for unigrams and bi-grams and take a list of top 7 unigrams and top 3 bigrams for each thread to be the list of keywords. Every sentence in a comment containing at least one of these keywords is taken as an argumentative sentence.
\par The next task is to identify claims and premises from these sentences. We hypothesize that the presence of some discourse connectives marks a sentence or clause as claim or premise; and when a discussion is going on a particular topic, these components show many similarities with each other throughout different comments. So, if there are $N$ comments in a thread comprised of a total $S_A$ argumentative sentences, and by searching for specific discourse markers we identify a set of claims $C'\subset C$ and premises $E'\subset E$ where $C$ and $E$ are the total collections of claims and premises respectively, then using some similarity measures we can differentiate $C-C'$ and $E-E'$ from $S_A$.
\par
\begin{figure}
    \centering
    \includegraphics[width=1.0\textwidth]{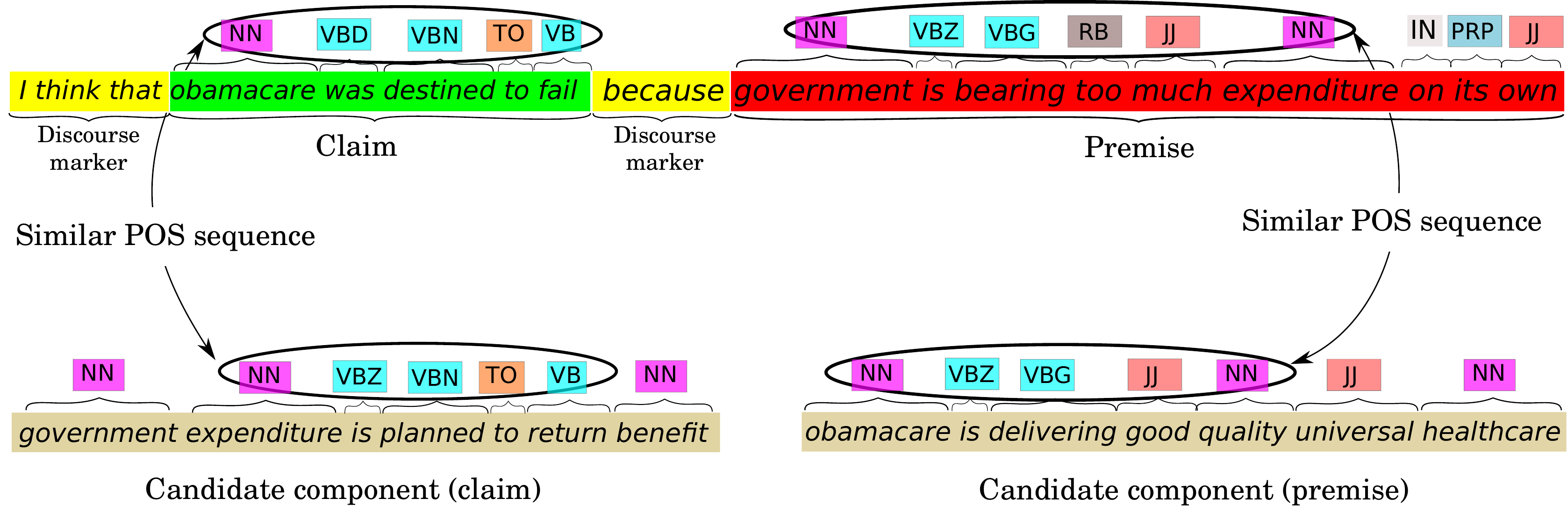}
    \caption{{\color{black}(Color online) Similarity between argument components; first sentence contains both claim and premise, identified by the presence of discourse markers ``\textit{I think that}'' and ``\textit{because}''. The two candidate segments shown below in the figure are undetected by the rules and need to be matched against the above components.}}
    \label{fig:arg_similarity}
\end{figure}

 For the primary detection of claims and premises, we employ the following rules:
\begin{itemize}
\item For a sentence of type ``\textit{I think that} \textbf{clause}" identify \textbf{clause} as a claim;
\item For a sentence of type ``\textit{In my opinion,} \textbf{clause}" identify \textbf{clause} as a claim;
\item For a sentence of type ``\textit{I argue that} \textbf{clause}" identify \textbf{clause} as a claim;
\item For a sentence of type ``\textbf{clause1} \textit{because} \textbf{clause2}" identify \textbf{clause1} as a claim and \textbf{clause2} as premise;
\item For a sentence of type ``\textbf{clause1} \textit{so} \textbf{clause2}" identify \textbf{clause2} as a claim and \textbf{clause1} as premise;
\item For a sentence of type ``\textbf{clause1} \textit{but} \textbf{clause2}" identify \textbf{clause2} as premise;
\end{itemize}
For each thread of discussions, we compute $C'$ and $E'$ using the above rules. To compute the rest of the components (undetected by the rules) we need to employ some similarity measure. \textcolor{black}{Similarity measures using vector space model (Euclidean distance, cosine similarity, etc.) are not a good option here; for a particular argumentation, claim and premise components may share similar words, resulting in higher similarity in vector space model. On the other hand, claim components for different argumentation may be less similar due to their semantic unrelatedness. We observe that similar components in an argument share similar syntactic and semantic structures. For example, in Fig.~\ref{fig:arg_similarity}, the claim and premise components identified by the rules have similar parts-of-speech (POS) sequences to candidate components. But if we look at the words used, alternate components have more similar words. The vector space model would identify ``\textit{obamacare was destined to fail}'' and ``\textit{obamacare is delivering good quality universal healthcare}'' as similar. In this example, the POS sequence explicitly identifies argument components; however, there can be more complex cases where POS and similar words together constitute similar sequences. For a complex linguistic task like argumentation understanding, it is important to observe words with their context -- how the words are being juxtaposed to construct the meaning. While representing words as fixed dimensional vectors, it is the temporal sequence as a whole that constructs their role in the argumentation and not the words as independent units. }

\textcolor{black}{For this, we treat each component $c_i\in C'$ and $e_i\in E'$ as a time-series constituted by words and their POS tags and calculate Dynamic Time Warping (DTW) distance between them.} DTW \cite{bellman1959adaptive,myers1980performance} is an efficient algorithm to compare time-series data. Given two time-series $P = \{p_1,p_2, ..., p_n\}$ and $Q = \{q_1,q_2, ..., q_m\}$, it constructs a  matrix $\mathbf{M}$ of size $n\times m$ recursively as follows:
\begin{equation}
\label{eq:dtw}
\mathbf{M}[i][j] = d(p_i,q_j) + min(\mathbf{M}[i-1][j],\mathbf{M}[i][j-1],\mathbf{M}[i-1][j-1])
\end{equation}
where $d(p_i,q_j)$ denotes the squared distance between $p_i$ and $q_j$.
$\mathbf{M}[n][m]$ then gives the cumulative sum of squared distances along the optimal path from (0,0) to (n,m). The DTW distance between $P$ and $Q$ is then given by $\sqrt{\mathbf{M}[n][m]}$.
\begin{figure}[h]
    \centering
    \includegraphics[width=0.7\textwidth]{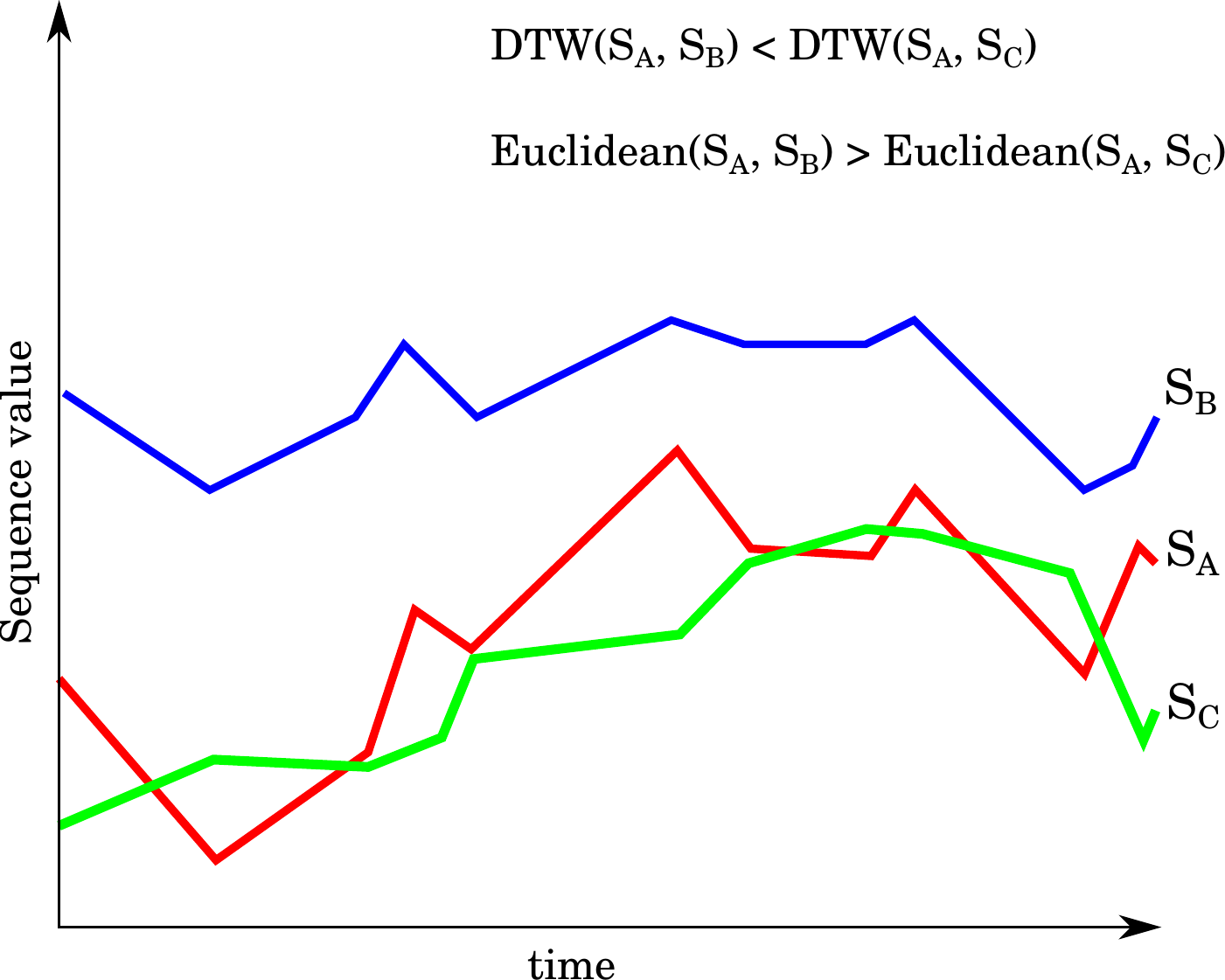}
    \caption{Three hypothetical 1-dimensional time series $S_A$, $S_B$ and $S_C$ are shown here; geometric shape of $S_A$ and $S_B$ are much similar compared to $S_C$, whereas values of $S_C$ are close to $S_A$. DTW distance and Euclidean distance between these sequences reflect the two types of similarity.}
    \label{fig:dtw_example}
\end{figure}
\par \textcolor{black}{DTW gives a geometric similarity measure between two n-dimensional sequences, as explained in Fig.~\ref{fig:dtw_example}.} We represent a text segment $T$ as a sequence of word vectors augmented with parts-of-speech (POS) tag and compute DTW distance between two text segments. The distance function $d(p_i,q_j)$ in Eq.~\ref{eq:dtw} is computed as squared Eucledian Distance between two POS-augmented word vectors.
\par Given $C'$ and $E'$ as sets of claims and premises respectively detected by the rules, and $S_A$ as set of sentences identified as argumentative, we can now proceed to classify $s\in S_A$ as claim, premise or none. We define two threshold distances, $d_c$ and $d_e$, denoting average DTW distance between text segments in $C'$ and $E'$ respectively, and $s_c$ and $s_e$ denoting standard deviation of inter-segment DTW distances in $C'$ and $E'$ respectively:
\begin{equation}
d_c = \frac{\sum_{c_i,c_j\in C'} DTW(c_i,c_j)}{N_c}
\end{equation}
\begin{equation}
d_e = \frac{\sum_{e_i,e_j\in E'} DTW(e_i,e_j)}{N_e}
\end{equation}
\begin{equation}
s_c = \sqrt[]{\frac{\sum_{c_i,c_j\in C'} (d_c-DTW(c_i,c_j))^2}{N_c}}
\end{equation}
\begin{equation}
s_e = \sqrt[]{\frac{\sum_{e_i,e_j\in E'} (d_e-DTW(e_i,e_j))^2}{N_e}}
\end{equation}
where $N_c$ and $N_e$ are the number of unique text segment pairs in $C'$ and $E'$ respectively. For any $s\in S_A$ if $\min_{c_i\in C'}{DTW(s,c_i)}\leq d_c$ then we decide $s$ to be a claim; if $\min_{e_i\in E'}{DTW(s,e_i)}\leq d_e$ then we decide $s$ to be an premise. If for any $s$ both of these conditions hold true to make a tie, we check whether
\begin{equation}
\label{ineq:tie}
 \frac{d_c-\min_{c_i\in C'}{DTW(s,c_i)}}{s_c}< \frac{d_e-\min_{e_i\in E'}{DTW(s,e_i)}}{s_e} 
\end{equation}
If the above inequality holds true, then $s$ is more probable to be in the premise set compared to the claim set, and vice versa otherwise.
\section{Experiment Setup}
\label{experiment}
In this section, we briefly discuss the dataset used for the experiments and experimental setups for training and testing our algorithms.
\subsection{Dataset}
We use the CMV dataset by Tan et al. \cite{tan2016winning} as our primary data. This dataset contains $18,363$ training threads of discussion over different topics, out of which in $25,09$ threads OP has awarded $\Delta$ to at least one chain of comments. Each thread is a tree of comments rooted at the OP comment, which starts the thread. We separate every chain from the thread by breadth-first tree traversal.
The length of the chains varies from 2 to 11. As there is no case of successful persuasion in chains of length 2, we take chains of length 3 to 11 for training and testing purposes. We take all such chains from the $25,09$ threads, totaling 38151 chains. With this, we take $17,048$ chains from $1,304$ threads where OP was never awarded a $\Delta$. This results in a total of $55,199$ chains, among which $7,370$ chains correspond to successful persuasion. For testing purposes, we used $15,304$ chains from the held out data in CMV, among these $2,239$ chains indicate successful persuasion.
\par For the argument extraction task, we manually annotated $20$ threads from CMV with claim and premise components. This results in $907$ comments annotated at word-level granularity for argument components. This leaves us with $1,564$ claim segments and $2,877$ premise segments. \textcolor{black}{Three expert annotators\footnote{The annotators were experts in computational linguistics and are of age between 25-35 years.} from linguistic background were told to identify argument components. We used the same annotation guidelines as Biran and Rambow \cite{biran2011identifying}. Cohen's $\kappa$ for the inter-annotator agreement was found to be $0.79$ for claims and $0.72$ for premises.}
\subsection{Word Embedding}
We generate word vectors from the CMV training and held-out set using Word2Vec \cite{mikolov2013distributed}. Five million sentences were used to train the skip-gram model for 500 epochs, to finally generate word embeddings of size 300. \textcolor{black}{We use these vectors to initialize the embedding layer in our persuasion model. Thus, the pre-trained word vectors are again trained for the task while learning to model persuasion.} For the argument extraction task, word vectors from the trained persuasion model are used. \textcolor{black}{As discussed in Sec.~\ref{sec:arg_ext}, each word was augmented with its parts-of-speech tag using Penn Treebank POS tagset \cite{marcus19building}. Out of 56 tags, our data used 43. Thus, for the semi-supervised argument extraction task, each word in a text segment is represented as a vector of length 343.} 
\subsection{Training Persuasion Detection Model}
To train the deep LSTM model explained in Sec.~\ref{sec:persuasion_model}, we use 80-20 training-validation split on the training set, with 5-fold cross-validation. Each fold was trained for 50 epochs, using \textit{Adam} optimization algorithm \cite{kingma2014adam}. \textcolor{black}{To handle the class imbalance in data, we used logarithmic class weighting on training data for persuasion identification; we compute weight as $w = \log \frac{n}{p}$, where $n$ and $p$ are the number of negative and positive samples in training data, respectively. Then Eq.~\ref{loss_2} is modified to, 
\begin{equation}
\label{loss_2}
\mathit{l_2} = -(wy\log p + (1-y)\log (1-p))
\end{equation}}
\subsection{Testing for Persuasion Prediction}
With the trained neural model for persuasion detection of complete chains, we set up another experiment for persuasion prediction on incomplete chains. In this case, we take chains of length 6 to 11 and make our model forecast whether persuasion will be successful while looking up to first 3 comments, first 4 comments and so on to the last comment.
\begin{figure}[!t]
        \centering
        \includegraphics[width=0.95\textwidth]{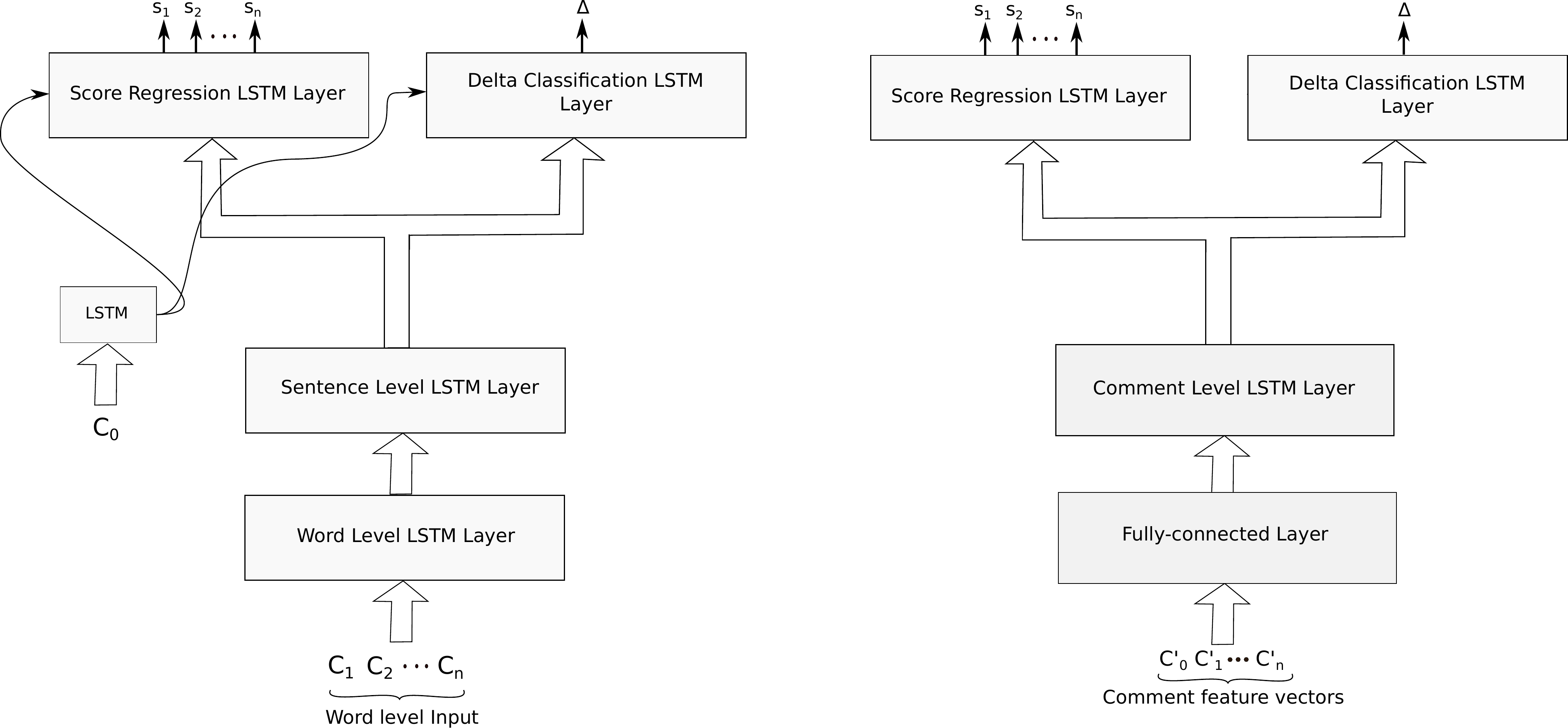}
        \caption{Architecture of hierarchical LSTM without attention (left) and LSTM with selected features (right). $C_i$ represents one hot encoding of words for $i^{th}$ comment. $C'_i$ represents feature vector for $i^{th}$ comment. For LSTM with selected features first comment is included in the chain input.}
        \label{fig:baseline}
    \end{figure}
\textcolor{black}{\subsection{Baselines for Persuasion Modeling}
We consider the following baseline algorithms for persuasion detection and comment score prediction to compare with the proposed method.
    \subsubsection{Hierarchical LSTM without Attention (HLSTM)}  
    We implement an LSTM-based architecture similar to ours but without employing the attention mechanism. We use the same set of pre-trained embeddings, optimizers, and training strategies. This model also performs joint prediction of both the tasks.
    \subsubsection{LSTM with Selected Features from Comments (LSTM-f)}
    We represent each comment text as a set of selected features. We use syntactic n-grams \cite{agarwal2009contextual,sidorov2014syntactic,sidorov2014should}, number of positive, negative and neutral polarity words computed using SentiWordNet \cite{baccianella2010sentiwordnet}, number of words, number of sentences and cumulative entropy of a comment token set $c$ as $H_c = \frac{1}{|T|}\sum_{t\in c}tf_t(\log |T| - \log tf_t)$ where $t$ is a unique term in $c$, $T$ is the set of all unique terms in the corpus (stopwords excluded and lemmatized), $tf_t$ is the frequency of $t$ in $c$. Each chain of comments is now a sequence of fixed length vectors, which is then processed by LSTM layers for score regression and $\Delta$-classification. The architectures of both of these two baseline models are illustrated in Fig.~\ref{fig:baseline}.
    \subsubsection{Support Vector Regression with Selected Features (SVR-f)} We implement a Support Vector Regression model with features computed for comments. This model is used only for the score regression task. We use the same set of features as in our LSTM with selected features model, i.e., syntactic n-gram, polarity word counts, number of words, number of sentences and cumulative entropy. We also use the FOG readability index \cite{gunning1969fog} of individual comments as an additional feature for this model.
\subsection{Baselines for Unsupervised Argument Extraction}
As already discussed, our focus is to tackle the argument mining problem with minimal supervision. As we extract a small subset of argument components using the rules already discussed and then use similarity matching to find the rest of the components, we implement the following baselines for similarity computation.
\subsubsection{Cosine Similarity} Given the seed component $s$ and candidate component $c$, we compute vectors $V_s$ and $V_c$ as TF-IDF weighted average of their word vectors respectively. Then $\alpha=\frac{V_s^\top V_c}{|V_s|\cdot|V_c|}$ provides the cosine similarity between $s$ and $c$.
\subsubsection{Word Mover's Distance} We compute semantic similarity between seed and candidate components using word mover's distance given by Kusner et al. \cite{kusner2015word}.
\subsubsection{KL Divergence} We compute the Kullback-Leibler divergence between L1 normalized bag-of-words vectors of seed and candidate components.
}

\begin{figure}[!t]
\centering
\includegraphics[width=0.8\textwidth]{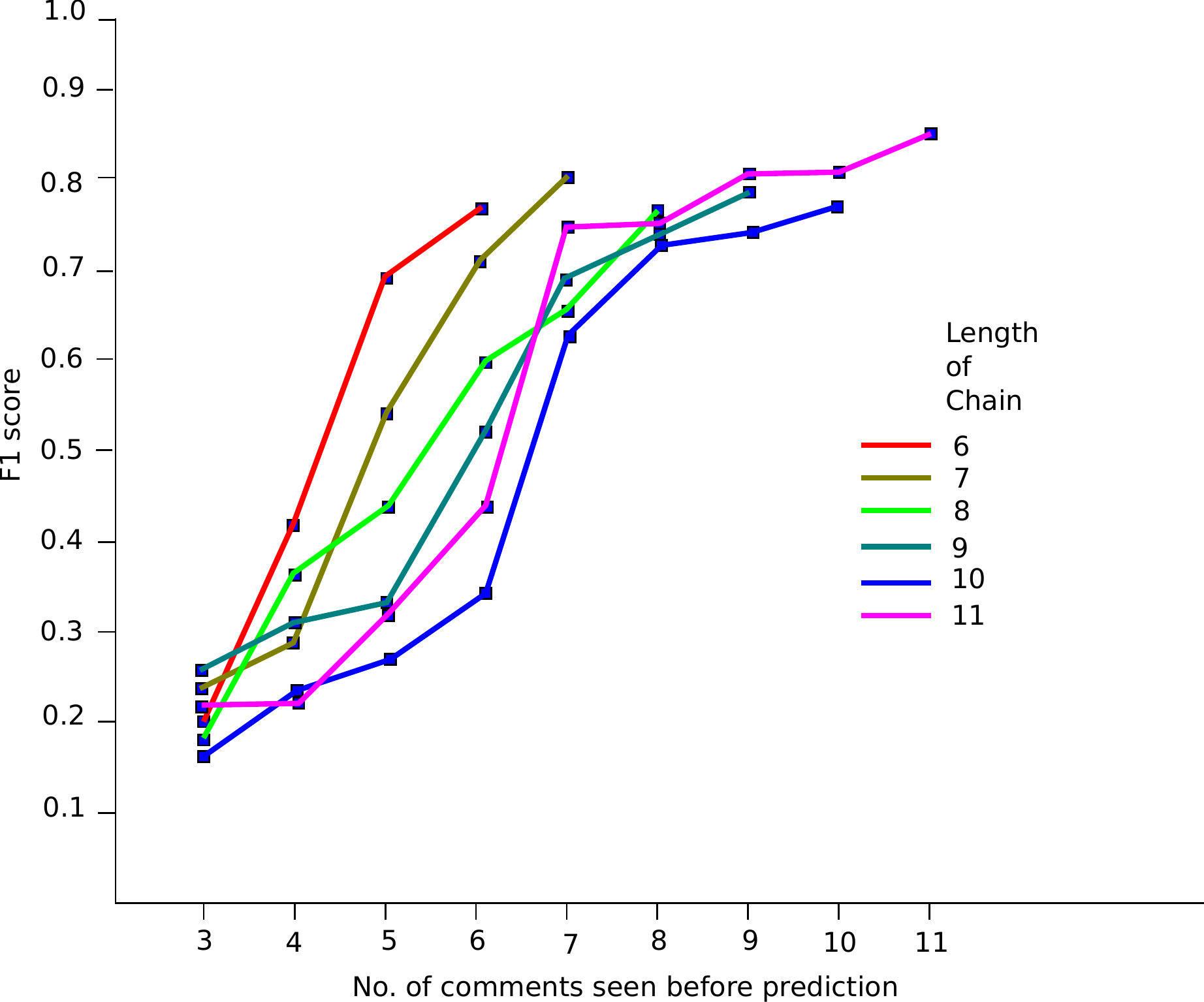}
\caption{(Color online) Persuasion prediction results for hierarchical LSTM with attention plotted against the number of comments observed prior to prediction.}
\label{fig:prediction}
\end{figure}

\section{Observations}
\label{sec:obs}
\begin{table}[!t]
    \centering
    \begin{tabular}{l|c|c|c|c}\hline
         Model & Precision & Recall & F1 score & AUC\\ \hline
         HLSTM-att & 0.85 & 0.73 & 0.79 & 0.71 \\
         HLSTM & 0.82 & 0.66 & 0.73 & 0.65 \\
         LSTM-f & 0.76 & 0.69 & 0.72 & 0.66 \\
         \hline
    \end{tabular}
    \caption{{\color{black}Performance of persuasion detection models.}}
    \label{tab:persuasion_allmodel}
\end{table}
Our persuasion detection model using hierarchical LSTM with attention clearly outperforms the baseline models with  \textbf{0.79} F1 score to classify $\Delta$ awarded chains. As we can see in Table~\ref{tab:persuasion_allmodel}, hierarchical LSTM models (with and without attention), which use word vectors as input have low recall value compared to the precision. With selected features, recall is actually higher than HLSTM without attention. Table~\ref{tab:chainwise_result} shows the evaluation results
\begin{table}[h]
\begin{center}
\begin{tabular}{c|c}
\hline Length of chain & F1 score \\ \hline
3 & 0.85 \\ 
4 & 0.73 \\ 
5 & 0.80\\ 
6 & 0.77\\ 
7 & 0.80\\ 
8  & 0.76\\
9  & 0.78\\
10  & 0.77\\
11  & 0.87\\

\hline
\end{tabular}
\end{center}
\caption{Chain-length wise performance for persuasion detection.}
\label{tab:chainwise_result}
\end{table}
for comment chains with different lengths.
\par \textcolor{black}{We train our HLSTM with attention model without weighing up positive classes to see how class imbalance affects the performance of the models. As we can see, recall for identifying positive persuasion (smaller class) drops heavily when the model is trained without class weight. There is also a significant increase in identifying true negatives. The class weight forces the model to penalize false negatives more than false positives; thereby restricting the bias of the model towards the dominant class (negative in this case). This explains the slight increase in recall for negative class and precision for positive class without weight.}
\begin{table}[]
\small
    \centering
    \begin{tabular}{l|c|c|c|c}
    \hline
         Class & \specialcell{Precision\\with weight} & \specialcell{Precision\\without weight} & \specialcell{Recall\\with weight} & \specialcell{Recall\\without weight}  \\
         \hline
         Positive & 0.81 & 0.82 & 0.76 & 0.61\\
         Negative & 0.95 & 0.91 & 0.92 & 0.90\\
         \hline
    \end{tabular}
    \caption{{\color{black}Class-wise performance of HLSTM with attention when trained with and without class weighting.}}
    \label{tab:class_weight}
\end{table}
\par Fig.~\ref{fig:prediction} shows prediction results observed for comment chains of different lengths (6-11) when the model forecasts persuasion success at different stages of the chain. 
We can readily see that there is a sharp increase in the performance over 5th to 7th comment in the chain.
\par \textcolor{black}{We transform the outputs of score regression to a ranking problem. We rank the comments according to the original karma score received and compare with the ranking from the scores predicted by our model.} We evaluate the performance for ranking for the top-most comment, the top three and the top five comments, presented in Table~\ref{tab:comment_ranking}.
\begin{table}[h]

\begin{subtable}[t]{0.48\textwidth}
\small
\centering
\begin{tabular}[t]{l|c|c|c}
\hline Rank & MAP & MRR & Kendall $\tau$\\ \hline
top & 0.87 & 0.83 & -- \\ 
top-3 & 0.74 & -- & 0.62 \\ 
top-5 & 0.67 & -- & 0.54\\ 

\hline
\end{tabular}
\vspace{16.5pt}
\caption{Ranking performance for HLSTM with attention for top most, top 3 and top 5 comments. }
\end{subtable}
\begin{subtable}[t]{0.48\textwidth}
\small
\begin{tabular}[t]{l|c|c|c}
\hline Model & top & top-3 & top-5\\ \hline
HLSTM-att & 0.87 & 0.74 & 0.67 \\ 
HLSTM & 0.85 & 0.73 & 0.64 \\ 
LSTM-F & 0.85 & 0.71 & 0.67\\ 
SVR & 0.83 & 0.70 & 0.63\\ 
\hline
\end{tabular}
\caption{Ranking performance of all the models; evaluation metric used is Mean Average Precision.}
\end{subtable}
\caption{Evaluation of ranking most influential comments.}
\label{tab:comment_ranking}
\end{table}
\begin{figure}[h]
    \centering
    \includegraphics[width=1.0\textwidth]{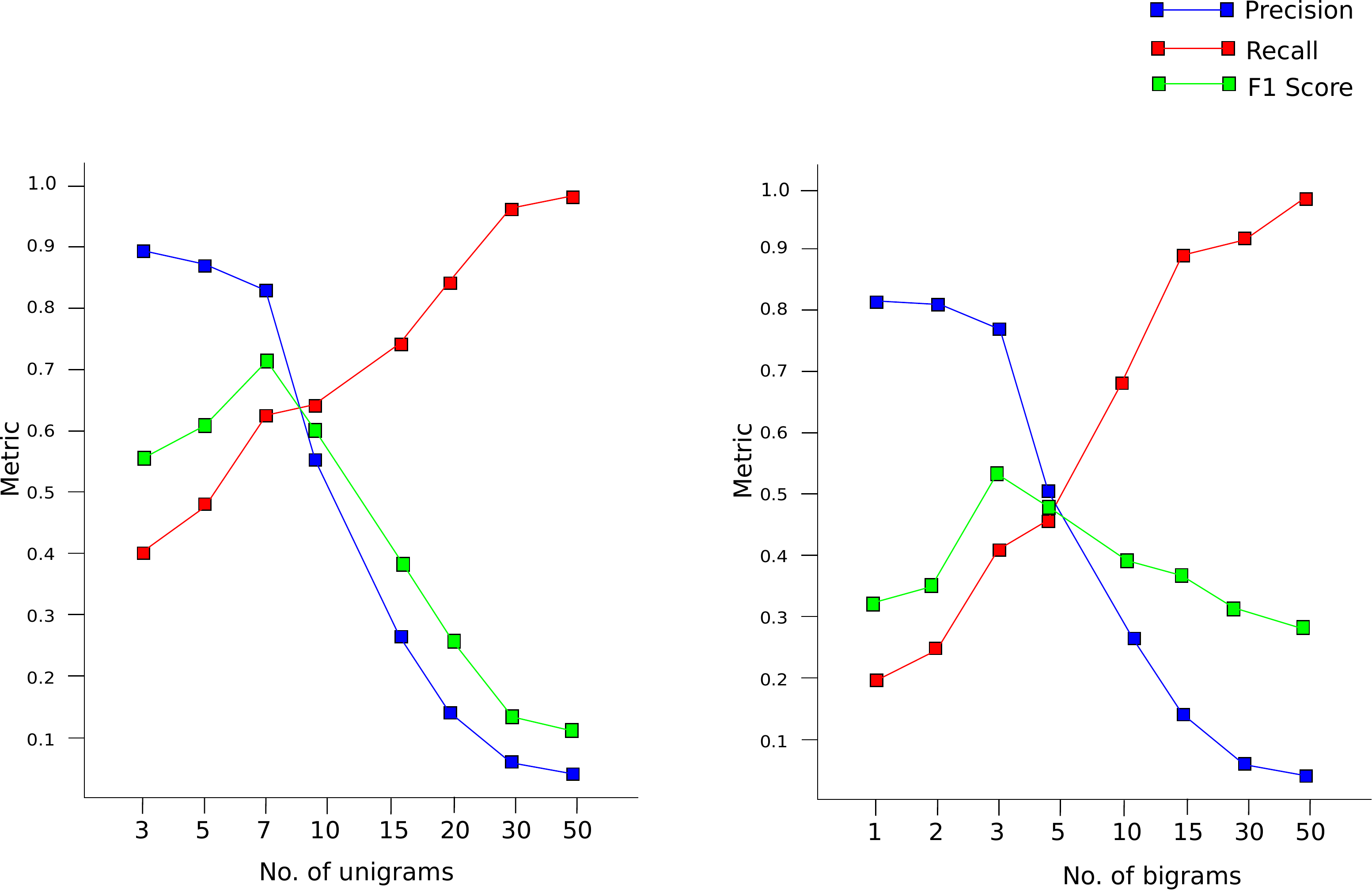}
    \caption{\textcolor{black}{Precision, recall and F1 score for argumentative sentence selection while using different number of top unigrams and bi-grams as keywords (used separately).}}
    \label{fig:keyword_selection}
\end{figure}

\par \textcolor{black}{We use different numbers of top (according to TF-IDF value) unigrams and bigrams in a discussion thread as keywords (mentioned in Sec.~\ref{sec:arg_ext}) to see how this selection effects the performance of argumentative sentence identification. We compute the performance of unigrams and bigrams separately. We notice in Fig.~\ref{fig:keyword_selection} that with the increase in the size of the keyword set, the precision value drops rapidly, and the recall value improves. Our choice of keywords (7 unigrams and 3 bigrams) is an optimal choice with the highest F1 score. While jointly using unigrams and bigrams for keywords, F1 score for argumentative sentence identification is $0.83$.}
\par Table~\ref{tab:argument_component} shows the performance of the semi-supervised argument component detection algorithm presented in Sec.~\ref{sec:arg_ext}. It is readily observable that, identifying claims by structural similarity is much easier compared to premises. Also, the high value of precision compared to lower recall indicates that the algorithm is able to detect specific types of argument components only. The linguistic rules employed can identify small subsets of argument segments with high precision on their own. 


\begin{table}[]
\small
    \centering
    \begin{tabular}{l|c|c|c|c|c|c}
\hline
\multirow{2}{*}{\specialcell{Argument\\ component}} & \multicolumn{2}{c|}{{ Precision}} & %
    \multicolumn{2}{c|}{{Recall}} & \multicolumn{2}{c}{{F1 score}}\\
\cline{2-7}
& { Rule} & { Rule+DTW} & { Rule} & { Rule+DTW} & { Rule} & { Rule+DTW}\\
\hline
Claim  & 0.97 & 0.68 & 0.22 & 0.41 & 0.36 & 0.51 \\ 
premise & 0.96 & 0.59 & 0.12 & 0.27 & 0.21 & 0.37 \\ 
\hline
\end{tabular}
    \caption{{\color{black}Evaluation of the semi-supervised model to identify argument components; performance of the full algorithm (linguistic rules + DTW similarity) and only linguistic rules is reported.}}
    \label{tab:argument_component}
\end{table}

\begin{table}[h]
\begin{center}
\begin{tabular}{l|c|c|c|c}
\hline Argument Segment & WMD & KLD & DTW & Cos-Dis\\ \hline   
Claim & 0.43 & 0.40 & 0.51 & 0.41\\
premise & 0.31 & 0.33 & 0.37 & 0.23\\
  
\hline
\end{tabular}
\end{center}
\caption{ Comparison of F1 scores for argument component identification with different similarity metrics.}
\label{tab:argument_compare}
\end{table}
As the comparison in Table~\ref{tab:argument_compare} shows, DTW proves to be a better measure compared to rest of the three. 

\par We extract the attention weights learned by our persuasion detection model for the argument component annotated threads. In Table~\ref{tab:attention_performance}, we compare the performance of the attention layer to identify argumentative sentences. We use multiple threshold weights above which sentences are taken.

\begin{figure}[!t]
\centering

\includegraphics[width=0.8\textwidth]{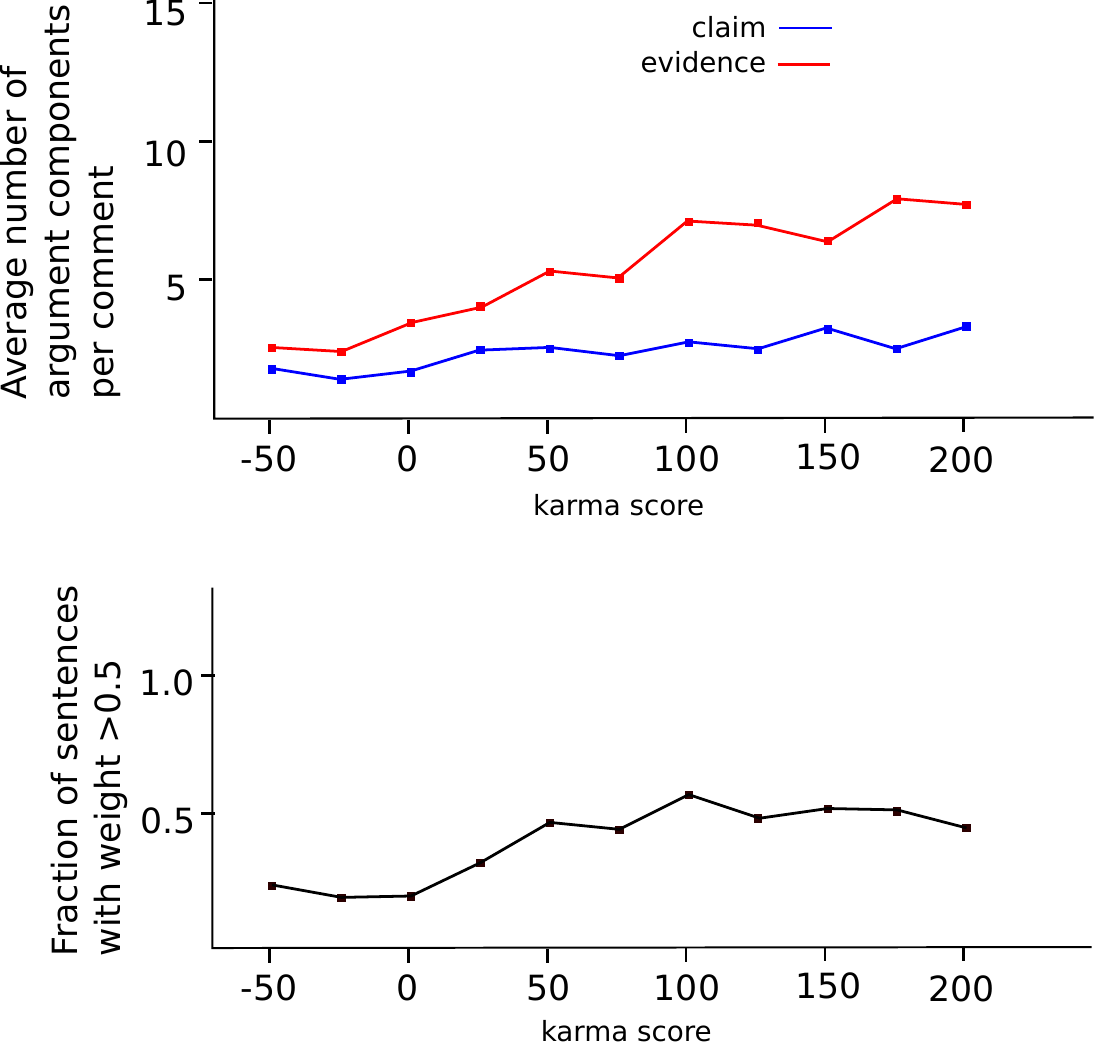}
\caption{Plots showing presence of argument components and attention weighting varying with karma score.}
\label{fig:com_arg}
\end{figure}

\begin{table}[h]
\begin{center}
\begin{tabular}{c|c|c|c}
\hline Sentences with weight$\geq$ & Precision & Recall & F1 score \\ \hline   
0.8 & 0.88 & 0.51 & 0.64\\ 
0.5 & 0.76 & 0.70 & 0.72\\
0.3 & 0.36 & 0.89 & 0.51\\
 
\hline
\end{tabular}
\end{center}
\caption{ Performance of attention layer to identify argumentative sentences. }
\label{tab:attention_performance}
\end{table}
\par It can be intriguing to check if the amount of argument components present in a comment is somehow correlated with its karma score. We plot the degree of presence of different argument components in a comment with its karma score. We take comments with karma score in a range between -50 to 200 as beyond that the distribution becomes too sparse. We also plot the fraction of sentences weighted $>$0.5 by the attention mechanism.  

As in Fig.~\ref{fig:com_arg}, we can see that the number of claim segments in a comment is not much related to its karma score; but for premises, a weak correlation can be observed. A similar mapping can be found in the fraction of sentences per comment which got weighted $>$0.5 by the attention mechanism while our deep LSTM model attempted to classify persuasion.

\section{Discussions}
\label{sec:discussions}
\begin{figure}[!t]
\centering
\includegraphics[width=0.8\textwidth]{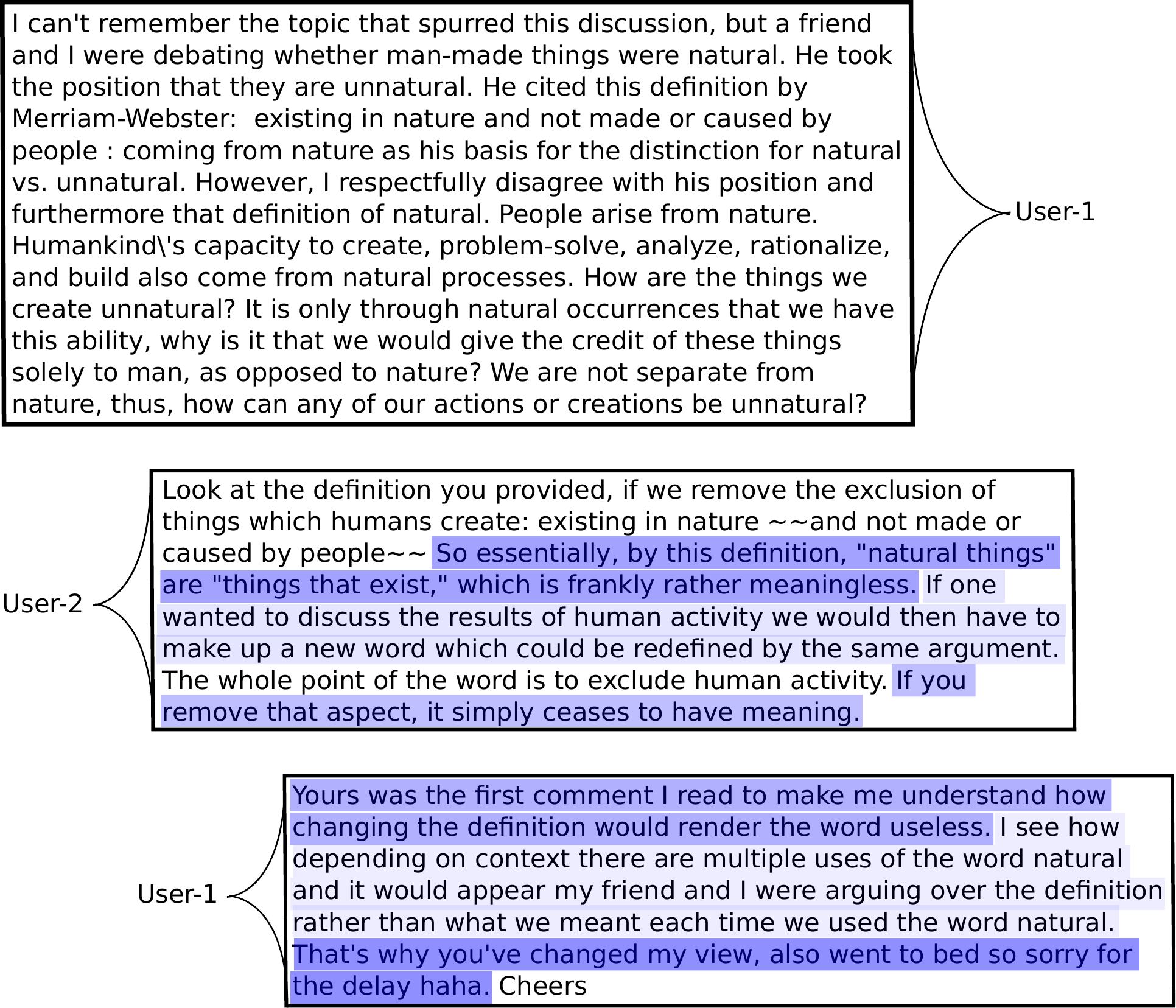}
\caption{Example of a chain with successful persuasion; shaded sentences represent sentences given weight$>$0.5 by the attention layer; darker shade means higher weight.}
\label{fig:atention_eg}
\end{figure}

We start by looking into the working of the attention mechanism in our model. In Fig.~\ref{fig:atention_eg}, a chain with three comments is presented, which achieved successful persuasion and our model detected it correctly. \verb|User-1| is OP here. We can see three characteristics of the weighted sentences in this example. In the last comment, the most weighted sentence is the one with the phrase \textit{``changed my view''}, indicating a marker of being persuaded. Intuitively, this sentence is important for only this reason, a linguistic cue denoting persuasion. If we check the most weighted sentences from the second comment and the second most weighted sentence from the last comment, they all focus on some definition being \textit{meaningless} or \textit{useless}. The focus of argumentation along the chain has been captured here. Lastly, all the weighted sentences shown here, except for the last one, are potential candidates of being argumentative sentences. A more detailed investigation of these cases may result in a better understanding of what type of features are being extracted by the deep LSTM model. We keep it as our future task.
\par In Table~\ref{tab:attention_performance} the performance of the attention mechanism weighting to identify argumentative sentences is presented. A quick check reveals that, when we take sentences with weight $>$0.8, 87\% of the sentences identified as argumentative contain premises, whereas only 39\% contain claims. Intuitively, this makes sense. It is the premises which make an argument more influential, not the claims. While identifying these components as a subtask of persuasion, it is only natural that the premises will be focused on.
\par As the results in persuasion detection and prediction suggests in Table~\ref{tab:chainwise_result} and Fig.~\ref{fig:prediction}, there are many cases where our model identified false positives or false negatives. Cases where OP used cues like \textit{you maybe correct but...}, the chain was erroneously detected as successfully persuaded. Also, where antagonistic topic words like \textit{Iran} and \textit{Israel} were used by OP and persuader interchangeably, our model could not identify the same view or opposing views correctly. The situation gets even more complicated when multiple users participate in a single chain, which is rare, but still a valid case. Incorporating target specific sentiment features may improve the results here.


\section{Conclusion and Future Work}
In this work, we proposed a deep LSTM model with an attention mechanism to jointly detect and predict persuasion in online discussions along with ranking comments by their persuasive potential. Our study revealed that our primary hypothesis of extracting argumentative sentences as a subtask of persuasion modeling holds true. The attention mechanism we employed could successfully focus on argumentative sentences in comments while learning persuasion. In fact, this attention mechanism could identify persuasive discourse markers. We proposed an algorithm for argument component detection with rules leveraging discourse connectives and Dynamic Time Warping distance as a structural similarity measure. We hypothesized that argument components in a common thread of discussion will maintain a similar structure, which holds true to an extent, as the results of testing our algorithm shows. 
\par Persuasion can be an intrinsic characteristic of influencer identification. Analysis of persuasive content in large networks can reveal a deeper understanding of online social interactions. Our work can be extended towards modeling argument diffusion over social networks. Incorporation of linguistic resources like sentiment lexicons, discourse treebanks along with deep learning can help achieve better results in both the problems we addressed -- persuasion modeling and argument extraction.
\section*{References}

\bibliography{mybibfile}

\end{document}